\documentclass[jgr]{agutex}
\usepackage{lineno}
\usepackage{graphicx}
\usepackage{color}
\usepackage{ulem}
\usepackage{amssymb}
\usepackage{amsthm}

\paperid{}
\journalid{2017}

\authorrunninghead{Lamy et al.}
\titlerunninghead{Uranus' aurorae past Equinox}

\begin{document}

\title{Uranus' aurorae past equinox}

\authors{L. Lamy\altaffilmark{1}, R. Prang\'e\altaffilmark{1}, K.~C.~Hansen\altaffilmark{2}, C.~Tao\altaffilmark{3}, S.~W.~H.~Cowley\altaffilmark{4}, T. Stallard\altaffilmark{4}, H. Melin\altaffilmark{4}, N. Achilleos\altaffilmark{5}, P. Guio\altaffilmark{5}, S.~V. Badman\altaffilmark{6}, T.~Kim\altaffilmark{7}, N.~Pogorelov\altaffilmark{7}}

\altaffiltext{1}{LESIA, Obs. de Paris, CNRS, UPMC, Univ. Paris Diderot, Meudon, France}
\altaffiltext{2}{Department of Atmospheric, Oceanic and Space Sciences, University of Michigan, Ann Arbor, Michigan, USA.}
\altaffiltext{3}{NICT, Tokyo, Japan}
\altaffiltext{4}{Department of Physics and Astronomy, University of Leicester, Leicester, UK}
\altaffiltext{5}{University College London, Mullard Space Science Laboratory, Dorking, UK}
\altaffiltext{6}{Lancaster University, Lancaster, UK}
\altaffiltext{7}{Center for Space Plasma and Aeronomic Research, University of Alabama, Huntsville, USA}

\begin{article}

\section{Abstract}

The aurorae of Uranus were recently detected in the far ultraviolet with the Hubble Space Telescope (HST) providing a new, so far unique, means to remotely study the asymmetric Uranian magnetosphere from Earth. We analyze here two new HST Uranus campaigns executed in Sept. 2012 and Nov. 2014 with different temporal coverage and under variable solar wind conditions numerically predicted by three different MHD codes. Overall, HST images taken with the Space Telescope Imaging Spectrograph reveal auroral emissions in three pairs of successive images (one pair acquired in 2012 and two in 2014), hence six additional auroral detections in total, including the most intense Uranian aurorae ever seen with HST. The detected emissions occur close the expected arrival of interplanetary shocks. They appear as extended spots at southern latitudes, rotating with the planet. They radiate 5-24~kR and 1.3-8.8~GW of ultraviolet emission from H$_2$, last for tens of minutes and vary on timescales down to a few seconds. Fitting the 2014 observations with model auroral ovals constrains the longitude of the southern (northern) magnetic pole to $104\pm26^\circ$ (284$\pm26^\circ$) in the Uranian Longitude System. We suggest that the Uranian near-equinoctial aurorae are pulsed cusp emissions possibly triggered by large-scale magnetospheric compressions.

\section{Introduction}

The Hubble Space Telescope (HST) recently succeeded in re-detecting the Far UltraViolet (FUV) aurorae of Uranus in 2011 and then in 1998 \citep{Lamy_GRL_12} (hereafter L12), long after their discovery by the UV Spectrometer (UVS) of Voyager 2 in 1986 \citep{Broadfoot_Science_86}. These detections included the first images of Uranus' aurorae and provided a new means to remotely investigate the poorly known magnetosphere of Uranus from Earth, awaiting for any future in situ exploration \citep{Arridge_EA_11}. This asymmetric magnetosphere has no equivalent in the solar system, with a spin axis close to the ecliptic plane, a 84-year revolution period which carried Uranus from Solstice in 1986 to Equinox in 2007, a fast spin period of 17.24$\pm0.01$~h and a $59^\circ$ tilt between the magnetic and the spin axes \citep{Ness_Science_86}. The geometry of the solar wind-magnetosphere interaction thus dramatically evolves over timescales ranging from a quarter of a rotation (hours) to seasons (decades).

\begin{figure*}
\centering
\noindent\includegraphics[width=35pc,angle=-0]{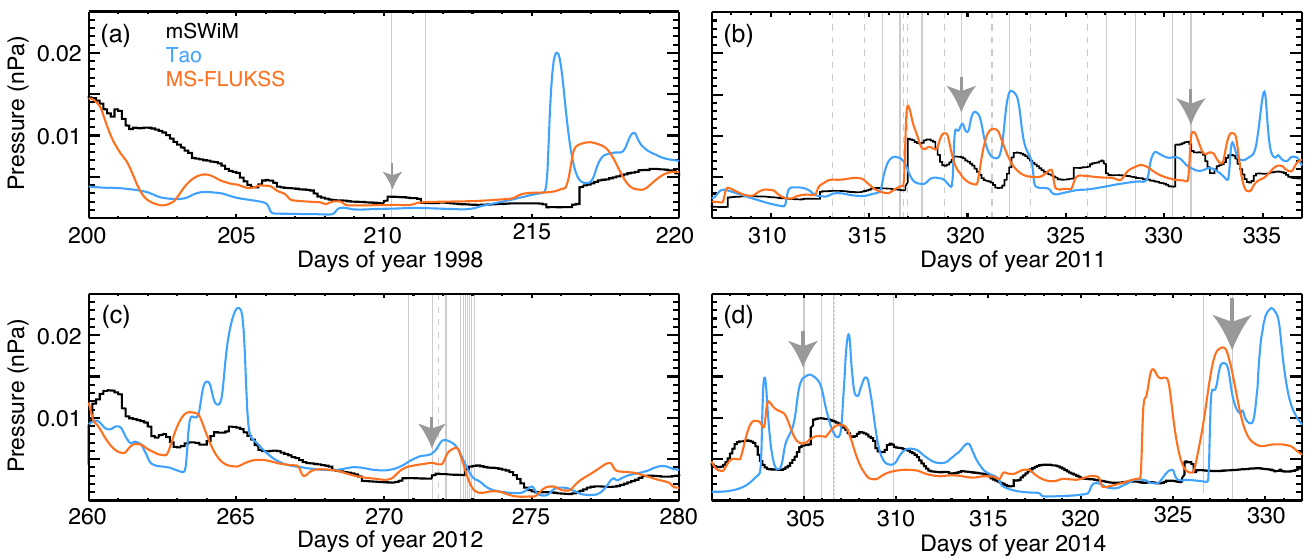}
\caption{Solar wind dynamic pressure at Uranus predicted by three MHD models (described in appendix \ref{models}) for the HST campaigns of (a) 1998, (b) 2011, (c) 2012 and (d) 2014. The uncertainty on pressure fronts is estimated to $\pm3$ days. Vertical gray lines mark the distribution of HST orbits using STIS (solid), ACS (dashed) and COS (dotted) instruments. Gray arrows indicate positive auroral detections with a size qualitatively proportional to their intensity.}
\label{fig1}
\end{figure*}

The 2011 HST observations were scheduled to sample the arrival at Uranus of a series of successive interplanetary shocks (displayed in Figure \ref{fig1}b), tracked through in situ solar wind measurements near Earth and numerically propagated to Uranus with an updated version of the Michigan Solar Wind Model (mSWiM), validated up to Saturn's orbit \citep{Zieger_JGR_08}. The observations acquired with the Space Telescope Imaging Spectrograph (STIS) yielded positive detections of auroral signal in two images (out of eight) analyzed by L12 and one spectrum studied by \citet{Barthelemy_Icarus_14}, and brought the first insights onto the Uranian magnetosphere near Equinox. The images revealed isolated auroral spots on 16 and 29 Nov. 2011 (gray arrows in Figure \ref{fig1}b), lasting for a few min, radiating a few kilo-Rayleighs (kR) over the observed FUV range. They were precisely colocated, rotationally phased in longitude and at $-10^\circ$ latitude. Their occurrence near times of predicted increases of solar wind dynamic pressure (up to 0.01~nPa) suggested that the solar wind could play a significant role in driving dayside auroral bursts. A STIS spectrum taken immediately after the STIS 29 Nov. 2011 image revealed auroral H$_2$ emission, radiating in average 650~R between 70~nm and 180~nm over the portion of the disc covered by the slit.

The re-analysis of STIS images of Uranus taken in 1998, in a configuration intermediate between Solstice and Equinox, yielded an additional detection during quiet solar wind conditions (gray arrow in Figure \ref{fig1}a). Although fainter and closer to the detection threshold than in 2011, the 1998 aurorae were seen in both hemispheres simultaneously and more spatially extended along ring-like structures reminiscent of partial auroral ovals. 


The emissions detected with HST contrasted with the Earth-like aurorae discovered by UVS at Solstice. The latter were clustered on the nightside, mainly around the southern magnetic pole along magnetotail longitudes, and radiated up to 3-7~GW in the H Ly$\alpha$ line and in the H$_2$ bands $\le116$~nm, $i.e.$ roughly twice as much over the full 70-180~nm H$_2$ range \citep{Herbert_JGR_94}. The variation of auroral characteristics along the Uranian orbit thus provides a diagnostic of the solar wind/magnetosphere interaction at very different timescales, which L12 assigned to changes of the magnetospheric configuration, through particle acceleration mechanisms yet to be identified. 

Two recent studies investigated possible origins of the observed auroral precipitations. \citet{Cowley_JGR_13} discussed the configuration of the Uranian magnetosphere at Equinox which inhibits the formation of a magnetotail. Under such conditions, the Uranian magnetosphere appears unable to drive bright, long-lasting auroral storms such as those observed at the Earth or Saturn induced by sudden magnetospheric compressions. \citet{Masters_JGR_14} modelled magnetopause reconnection at both Solstice and Equinox using Voyager 2 solar wind parameters and concluded that dayside reconnection is in general less favorable at Uranus than at inner planets, at Equinox than at Solstice, and predicted highly dynamic reconnection sites. 


In this article, we analyze two new HST campaigns executed in Sept. 2012 and Nov. 2014 with different temporal coverage and under variable solar wind conditions (section \ref{obs}). The images provide six additional detections of Uranus aurorae, whose properties display both similarities and differences with those of auroral emissions detected in 2011 (section \ref{results}). All Uranian aurorae seen by HST are then discussed together to investigate any possible control by the solar wind and/or by the planetary rotation (section \ref{discus}).

\begin{table*}
\center
\small{
\begin{tabular}{c|c|c|c|c|c|c|c|c}
  Date (Earth time) & Dataset & Filter & Exposure & CML & Latitude & Longitude & Peak brightness & Total Power \\
  \hline
  1998-07-29 06:07:43 UT & o4wt01t0q & 25MAMA & 1020s & $180^\circ$ &  $35\pm35^\circ$ & $93\pm23^\circ$ & 4 kR & $ - $ \\ 
  \hline
  2011-11-16 15:32:10 UT & obrx10p0q & 25MAMA & 1020s & $338^\circ$ &  $11\pm3^\circ$ & $49\pm5^\circ$ & 11 kR & $2.0\pm0.8$ GW \\ 
  2011-11-29 02:09:24 UT & obrx18hbq & 25MAMA & 1020s & $93^\circ$ &  $9\pm3^\circ$ & $55\pm3^\circ$ & 10 kR & $2.4\pm0.8$ GW \\ 
  \hline
  2012-09-27 15:00:19 UT & obz501dgq & 25MAMA & 1250s & $296^\circ$ &  $-50\pm3^\circ$ & $297\pm11^\circ$ & 5 kR & $1.9\pm1.3$ GW \\ 
  2012-09-27 15:27:07 UT & obz501diq & F25SrF$_2$ & 820s & $304^\circ$ &  $-49\pm4^\circ$ &  $294\pm11^\circ$ & 15 kR & $2.2\pm1.8$ GW \\
  \hline
  2014-11-01 23:57:33 UT & ocpl02nzq & 25MAMA & 1231s & $111^\circ$ & $-40\pm4^\circ$ & $105\pm7^\circ$ & 6 kR &  $1.3\pm1.0$ GW\\
   2014-11-02 00:26:11 UT & ocpl02o6q & F25SrF$_2$ & 900s & $120^\circ$ & $-38\pm4^\circ$ & $105\pm13^\circ$ &15 kR & $-$  \\  
   2014-11-14 08:34:22 UT & ocpl07ckq & 25MAMA & 757s & $155^\circ$ & $-44\pm9^\circ$ &  $105\pm15^\circ$& 17 kR & $5.9\pm1.4$ GW\\
    2014-11-14  09:04:00 UT & ocpl07cmq & F25SrF$_2$ & 900s & $165^\circ$ & $-42\pm10^\circ$  &  $115\pm10^\circ$ & 24 kR & $8.8\pm1.8$ GW\\
\end{tabular}
}
\caption{(Columns 1 to 5) HST observing parameters at mid-exposure. (Columns 6 to 9) Properties of auroral emissions detected by HST in 1998, 2011, 2012 and 2014.}

\label{tab}
\end{table*}

\section{Dataset}
\label{obs}

\subsection{HST observations}

Following the Nov. 2011 HST campaign, two subsequent HST programs were executed in Sept. 2012 and Nov. 2014, while Uranus gradually moved away from the 2007 Equinox. These two programs consisted of a total of 19 HST visits, each one lasting 1 orbit, which mainly used the Space Telescope Imaging Spectrograph (STIS, 17 orbits) but also the Advanced Camera for Surveys (ACS, 1 orbit) and the Cosmic Origin Spectrograph (COS, 1 orbit) \footnote{http://www.stsci.edu/hst/HST\_overview/instruments}. All the STIS and COS observations were acquired with the time-tag mode, which provides the arrival time of photons recorded on the MAMA detector at a 125~microsec time resolution. In this article, we analyze the STIS data obtained along 13 imaging orbits. We left aside ACS images which, as in L12, did not bring positive results. STIS spectra were already analyzed by \citep{Barthelemy_Icarus_14}, while the analysis of COS data is beyond the scope of this study. Each STIS imaging orbit was made of a pair of consecutive images taken with the Far-UV MAMA (Multi-Anode Microchannel Array) detector using the clear filter 25MAMA (137~nm central wavelength, 32~nm FWHM) which spans H$_2$ bands and H Ly-$\alpha$, and the Strontium Fluoride filter F25SrF$_2$ (148~nm central wavelength, 28~nm FWHM) which rejects wavelengths shortward of 128~nm, including H Ly-$\alpha$. 

The 2012 program was aimed at carefully sampling the rotational dynamics of auroral processes in order to assess the influence of rotation on the magnetosphere/solar wind interaction. The observations included 7 STIS imaging orbits spread from 27 to 29 Sept. 2012 over three consecutive planetary rotations, hence providing an excellent longitudinal coverage. This interval matched a modest increase of solar wind dynamic pressure (Figure \ref{fig1}c).

The main goal of the 2014 program, obtained with director's discretionary time, was to track the auroral response to two episodes of powerful interplanetary shocks characterized by large fronts of dynamic pressure at Uranus (Figure \ref{fig1}d) up to or beyond 0.02~nPa (depending on the solar wind model, see section \ref{models}), twice as large as in 2011 and thus the largest ever sampled by both HST and Voyager 2. The observations included 6 STIS imaging orbits distributed from 1 to 5 Nov. and from 22 to 24 Nov. 

\subsection{Image processing}

The data were processed exactly as in L12 with the simple, robust two-steps pipeline described below.
 

The STIS images were calibrated through the Space Telescope Science Institute pipeline and corrected for any geocoronal contamination, by subtracting to all pixels a constant offset intensity estimated beyond the disc. Indeed, F25MAMA images are highly sensitive to contamination at H Ly-$\alpha$ and the oxygen OI 130.4 nm multiplet, but even F25SrF$_2$ images can be affected by strong oxygen lines. The level of contamination was variable with time, resulting in a variable background level of STIS exposures. We then subtracted to each image an empirical model of disc background of solar reflected emission. This background model was built from a median image, derived separately for 25MAMA and F25SrF$_2$ filters and for each HST campaign, before to be fitted to and subtracted from each individual image. 

Although some of the images used to build our empirical background possibly include the auroral emissions we are looking for, the derived model is generally excellent, as the location of auroral spots far from the rotational poles together with their short lifetime renders it a priori unlikely to observe auroral signal exactly at the same position across the planetary disc in different images. This was a posteriori confirmed by the different location of detected auroral signal presented in section \ref{results}. The empirical background models were built for the 2012 and 2014 campaigns from a set of 7 and 6 images taken in each filter, respectively. The statistics was thus fair, but unsufficient to smooth out spatial inhomogeneities. 

Therefore, we also used an alternate numerical background model of background built with Minnaert functions \citep{Vincent_Icarus_00} fitted to the disc emission of each image and convolved by the STIS point spread function. This model, although less physical, is smooth and well suited to track isolated auroral features. Hereafter, we display images processed with the empirical background, but we required auroral signatures to be detected with both kinds of background models to be considered as positive detections.

Each background-subtracted image was then smoothed over a $5\times5$ pixels averaging filter to increase the signal-to-noise ratio (SNR). This choice, already used by L12, was checked by varying the size of the averaging filter and found to provide the best compromise between increasing the SNR and preserving the spatial resolution.

The processed images in counts were ultimately transposed into physical units of kR and GW of unabsorbed H$_2$ emission over 70-180~nm by using the conversion factors described in \citep{Gustin_JGR_12}. This enables one to compare brightnesses derived with different filters and more largely with different instrumentation.

\subsection{Solar wind models}
\label{models}

In L12, we used solar wind parameters at Uranus numerically propagated from the Earth orbit out to Uranus by one single MHD model, namely the Michigan Solar Wind Model (mSWiM) \citep{Zieger_JGR_08}. In the present study, we used the results of three different codes : mSWiM (1D), the Tao model (1D) \citep{Tao_JGR_05} and the Multi-scale Fluid-kinetic Simulation Suite (3D, MS-FLUKSS) \citep{Pogorelov_14}, all using near-Earth solar wind in situ observations provided by NASA/GSFC's OMNI 1h averaged data set through OMNIWeb \citep{King_JGR_05}. The results of these models are displayed by black, blue and orange lines in Figure \ref{fig1}, respectively. They are described in more details in appendix \ref{app_sw} by historical order of use and compared to infer their limitations. Overall, we estimate a typical uncertainty of $\pm3$ days on the dynamic pressure fronts at Uranus. 

As only MS-FLUKKS has been validated yet in the outer heliosphere by the comparison of predicted parameters with in situ plasma measurements of Ulysses, Voyager and New Horizons missions \citep{Kim_ApJ_16}, the MS-FLUKKS results (orange lines in Figures \ref{fig1} and \ref{figS1}) are hereafter taken as a primary reference to which the mSWiM and Tao results are compared.

\begin{figure*}
\centering
\noindent\includegraphics[width=35pc]{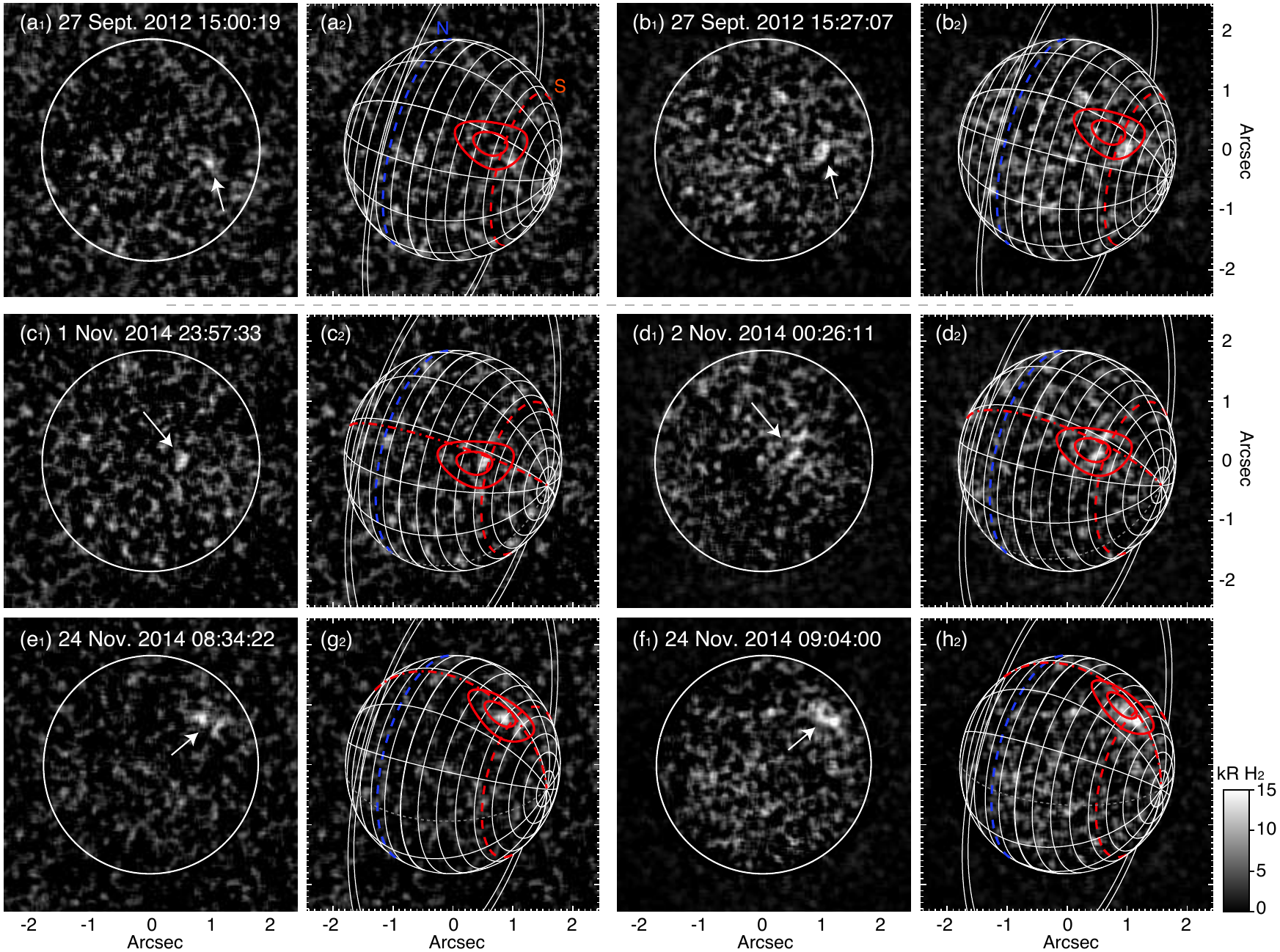}[!h]
\caption{HST/STIS images acquired on 27 Sept. 2012 (a$_1$-b$_1$), 1-2 and 24 Nov. 2014 (c$_1$-f$_1$) and replicated with grids of planetocentric coordinates (a$_2$-f$_2$). Images were acquired with the 25MAMA (first column) and the F25SrF$_2$ (third column) filters and processed as described in the main text. They are displayed in kR of unabsorbed H$_2$ emission over 70-180~nm. The observing times are in Earth UT. White arrows indicate spatially extended bright spots above the detection threshold. The planetary configurations are corrected for light time travel ($\sim$2.7 hours). The dotted grey meridian marks the 0$^\circ$ ULS longitude. The red and blue dashed parallels (dotted-dashed meridians) mark the latitude (longitude) of the southern and northern magnetic poles, respectively. Model southern auroral ovals fitted to the data are displayed by pairs of solid red lines (see main text). The conjugate model northern auroral oval, shifted by $180^\circ$ longitude, is not visible.}
\label{fig2}
\end{figure*}

\section{Average properties of auroral structures}
\label{results}

Simple criteria were used to identify auroral signatures : the emission region must reach or extend beyond a $4\times4$ pixels box with intensities per pixel exceeding 3 standard deviations ($\sigma$) above the background level. This is intended to discard isolated bright pixels. Inspection of all STIS images revealed six positive detections (out of twenty-six exposures, hence detections in roughly a quarter of exposures, strikingly similar to L12) displayed in Figure \ref{fig2} a$_1$-f$_1$ (and replicated in Figure \ref{fig2} a$_2$-f$_2$ with grids of planetocentric coordinates) and indicated by white arrows. These detections appear in three pairs of consecutive images taken on 27 Sept. 2012, 1-2 and 24 Nov. 2014. The corresponding observing parameters are indicated in columns 1-5 of Table \ref{tab}, which also includes the previous detections analyzed by L12 for comparison purposes. The peak intensity exceeded the 5$\sigma$ level in images c$_1$ and f$_1$, with $\sigma=2.5$~kR of H$_2$ in average. The acquisition of STIS images in pairs further strengthens these detections since the auroral signal is seen to persist from one image to the next and to rotate with the planet. This motion is consistent with the expected $8-9^\circ$ longitudinal shift derived from the CML difference between two consecutive exposures. 

Hereafter, longitudes refer to the Uranian Longitude System (ULS) \citep{Ness_Science_86}. ULS longitudes are built from IAU-defined longitudes, both increasing with time, by referencing the $168.46^\circ$ sub-Voyager 2 IAU longitude on 24 Jan. 1986 to $302^\circ$ according to the ULS definition. Absolute longitudes cannot be determined any more as the reference has been lost, owing to the large uncertainty on the rotation period. From 24 Jan. 1986 to 24 Nov. 2014, the planet rotated $14660.3\pm8.5$ times. In the ULS system, latitude is measured positively from the equator toward the rotation axis and the northern and southern magnetic poles lie at $+15.2^\circ$ and $-44.2^\circ$, respectively.

\subsection{Morphology}

These new auroral features display both strong similarities to and some differences from those detected in 2011. They appear as isolated spots, as in 2011, but with a larger spatial extent of up to several tens of pixels (1 pixel $\sim$ 340~km). These emissions all lie in the southern hemisphere, nearly at the southern magnetic pole latitude, while the 2011 aurorae appeared closer to northern polar latitudes. Columns 6-7 of Table \ref{tab} provide the coordinates of the auroral peak and its spatial extent at half maximum, assuming an auroral altitude at 1100km above the 1-bar level. This altitude is taken to be the same as for Saturn's aurorae and is consistent with early models of peak auroral energy deposition at Uranus \citep{Waite_JGR_88}.

As noted above, the auroral spots appear to persist and rotate with the planet during each pair of consecutive images. Quantitately, Table \ref{tab} shows that the peak emission on 27 Sept. 2012 and 1-2 Nov. 2014 did not vary by more than $2^\circ$ in latitude and $3^\circ$ in longitude, well within the extent of the auroral region. This suggests a single active region fixed in longitude. In contrast, on 24 Nov. 2014, the peak emission remains at constant latitude but shifts by $11^\circ$ in longitude. This compares with the larger size of the auroral region itself whose morphology (as well as intensity and dynamics, discussed below) significantly evolves from the first image to the second. 

Interestingly, the aurorae seen on 1-2 and 24 Nov. 2014, 22 days apart, appear at the same latitude and longitude. This indicates that, assuming an arbitrary southern auroral oval of constant size, the same portion of it was activated for different CML, as already observed in the north on 16 and 29 Nov. 2011, 13 days apart. The 27 Sept. 2012 aurorae were activated $10^\circ$ southward of the 2014 emissions, and at longitudes which cannot be compared to those of 2014 due to the large uncertainty in the ULS system ($\pm106^\circ$ per year).


\subsection{Energetics}

Figure \ref{fig2} displays images in kR of unabsorbed H$_2$ emission over 70-180~nm. A supplementary 16\% average contribution of H Ly$\alpha$ \citep{Broadfoot_Science_86} may be added to obtain an exhaustive estimate of the total flux radiated by H and H$_2$. Column 8 of Table \ref{tab} lists the H$_2$ auroral peak brightnesses, for the 2012 and 2014 campaigns but also for the 1998 and 2011 ones. These generally lie within a range of 5-15~kR. An exceptionally high value of 17-24kR was reached on 24 Nov. 2014. We note that, within each pair of observations, the second image systematically displayed a brighter signal. We attribute these changes to intrinsic auroral variability as the active region is clearly seen to simultaneously extend and brighten in each case. The brightnesses discussed above are roughly consistent with the few kR estimated by L12 for the 2011 auroral spots in the observed 25MAMA range, and they strikingly compare to (and in the case of 24 Nov. 2014 emissions even significantly exceed) the 9~kR of H$_2$ emission derived from Voyager 2/UVS measurements of southern nightside aurorae. Uranus aurorae are much less bright than Jupiter's but compare well with the average 10~kR of Saturn's aurorae (e.g. \citep[and references therein]{Lamy_JGR_13}).

To estimate the total radiated power, we derived the total number of counts per second within a constant radius circle encompassing the auroral signals (17 pixels $\sim5800$~km) . This size was chosen by fitting the largest spot in figure \ref{fig2}f$_1$ and then applied to all the images for the sake of consistency (except for the 1998 observation which displayed auroral features of different shape and wider than 17 pixels). Values were then converted into total H$_2$ power as described in section \ref{obs}. The results are provided in column 9 of Table \ref{tab} (except for figure \ref{fig2}d$_1$ which was contaminated by an irregular glow on the detector preventing any reliable power estimate). The large associated uncertainty has been estimated separately for each image. This uncertainty divides into $\sim$1/3 of Poisson noise and $\sim$2/3 of error on the background. The resulting power ranges from 1.3$\pm1.0$~GW on 1 Nov. 2014 to 8.8$\pm1.8$~GW on 24 Nov. 2014. Assuming the canonical 10\% efficiency between precipitated and radiated power, the precipitated power ranges from 13 to 88 GW. The radiated powers again compare with (as for brightnesses) but here do not exceed the $\sim$6-14~GW inferred from Voyager 2/UVS measurements of southern nightside aurorae. This likely results from emissions less spatially extended near equinox than at solstice. Similarly, such values remain lower than the usual power radiated by Saturn's aurorae, which extend along wide, circumpolar ovals.


\begin{figure*}[!h]
\centering
\noindent\includegraphics[width=26pc]{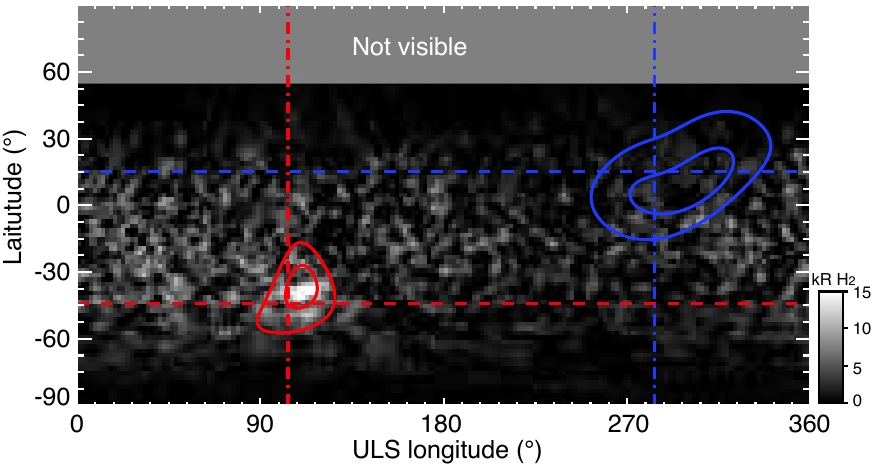}
\caption{Composite cylindrical projection built from the 12 STIS processed images of Uranus obtained in Nov. 2014. The top white region indicates latitudes which could not be sampled. The average H$_2$ brightness was derived in $2^\circ\times2^\circ$ bins. Uranocentric coordinates are taken at 1100~km above the 1-bar level. Red and blue pairs of solid lines indicate southern and northern model auroral ovals calculated with the AH5 model. Their outer and inner boundaries map the footprint of field lines whose apex reach 5 and 20~R$_U$ respectively. The red and blue horizontal dashed parallels indicate the latitude of magnetic poles. The red and blue vertical dotted-dashed meridians indicate the best-fit longitude of magnetic poles, namely $104\pm26^\circ$ (284$\pm26^\circ$) for the southern (northern) pole.}
\label{fig3}
\end{figure*}
 
\subsection{Dynamics}

The auroral dynamics appears to differ slightly from what was observed in 2011. The latter were seen to vary on timescales of minutes. Here, the auroral signatures persist over longer intervals, covered by two consecutive images. From the delay between the mid-exposure times of consecutive images, the active region lasts for at least $\sim$17, 18 and 13 min on 27 Sept. 2012, 1-2 and 24 Nov. 2014, respectively. Within these active periods, variations and recurrences can be observed on much shorter timescales.


To investigate this dynamics in more details, we performed a time-tag analysis of the brightest auroral features seen on 24 Nov. 2014. The time-tag mode enables us to process the data at the desired time resolution and to build time series of the counts recorded in a specific region of the detector. The auroral signal detected on 24 Nov. 2014 was sufficiently high to motivate the analysis of its temporal dynamics over the exposure time of the two images displayed in figures 2e$_1$ (clear filter 25MAMA) and 2f$_1$ (filter F25SrF$_2$). As reminded in table \ref{tab}, these images were acquired successively at 08$:$34$:22$ and 09$:$04$:$00~ÊUT (Earth time) and integrated over 757~s and 900~s respectively. The lower effective integration time of the former 25MAMA image (compared to other F25SrF$_2$ or 25MAMA images) is due to an unusually high count rate dominated by geocoronal contamination which, in turn, saturated the onboard buffer memory before the data could be transferred, resulting in several significant data gaps. 

\begin{figure*}
\centering
\noindent\includegraphics[width=35pc]{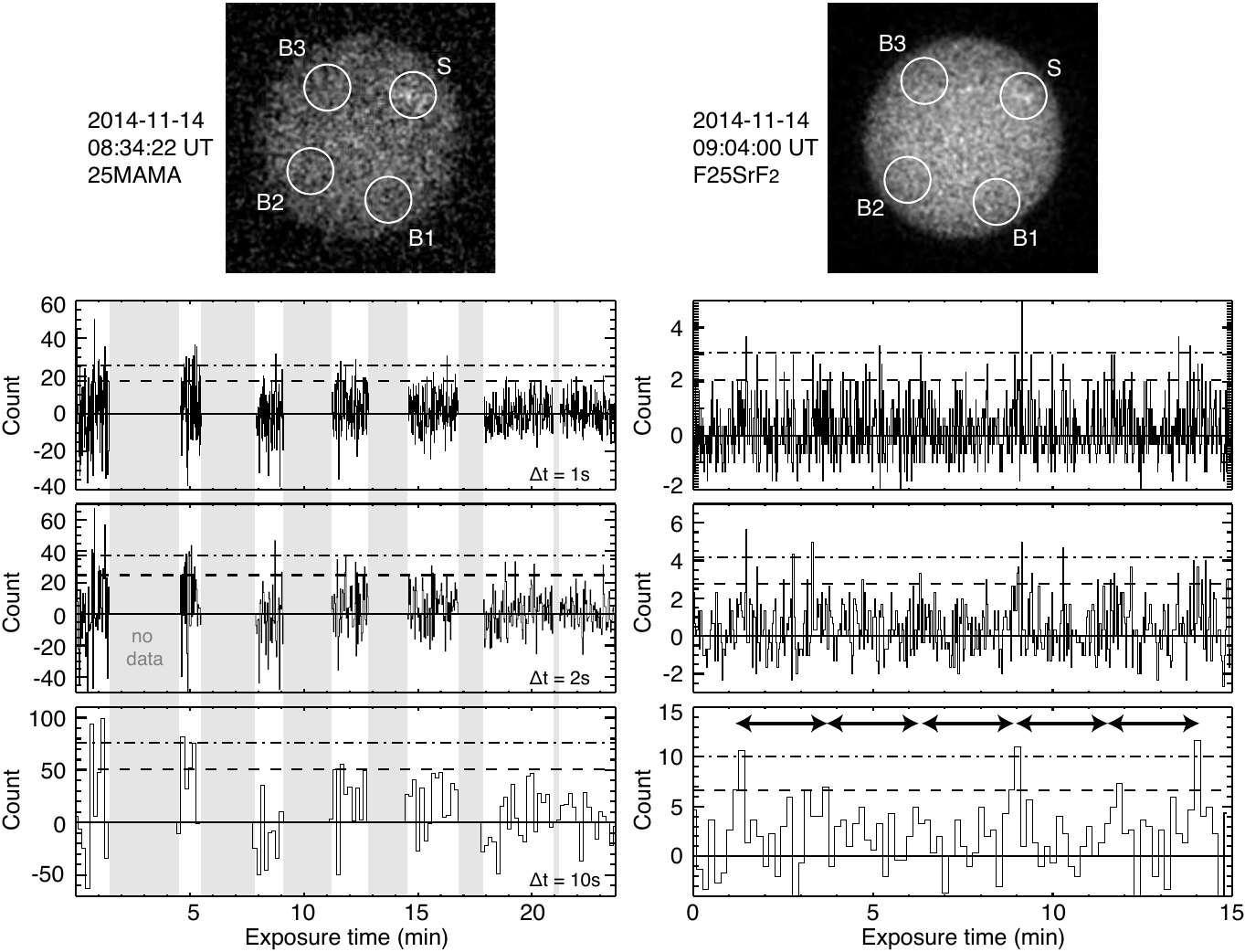}
\caption{Consecutive images of Uranus acquired on 24 Nov. 2014 with the 25MAMA and F25SrF$_2$ filters. White circles define discs mapping regions with and without auroral emission. The disc labelled S surrounds the auroral region and served to determine the signal count rate. The discs labelled B$1$, B$_2$ and B$_3$ surround background regions at similar solar zenithal angles, B$_1$ being additionally chosen at the same latitude as S. The signal averaged over discs B$_1$-B$_3$ served to determine a mean background count rate. The three histograms below each image display time series of the difference between the signal and the background count rate with different time resolution, namely 1~s, 2~s and 10~s from top to bottom respectively. Horizontal dashed and dashed-dotted lines indicate the 2 and 3$\sigma$ level above the background.}
\label{fig4}
\end{figure*}

Figure \ref{fig4} replicates figures \ref{fig2}e$_1$-f$_1$. On top of each image, four 17 pixels wide white circles are drawn, defining four discs over each of which a count rate was derived. A disc surrounding the auroral emission region (labelled S) was first used to determine the signal count rate. The three other discs (labelled B$_1$ to B$_3$) were chosen out of the auroral region at similar solar zenithal angles across the planet, with B$_1$ being additionally chosen at the same latitude as S. The signal averaged over discs B$_1$-B$_3$ served to determine a background count rate with a low noise. Time series of the difference between the signal and the background count rate are displayed below each image of Figure \ref{fig4} with three different temporal resolutions : 1~s, 2~s and 10~s from top to bottom respectively. Hereafter, we pay specific attention to episodes which reached or exceeded 2 or 3 standard deviations $\sigma$ above the background level (indicated by horizontal dashed and dashed-dotted lines respectively), although the $\sigma$ reference may be slightly over-estimated due to the presence of auroral emission.

Although the 25MAMA image was built over discontinuous intervals, the 10~s integrated histogram clearly displays 4 peaks in excess of 3$\sigma$ during the first minute of integration. The 10~s integrated histogram corresponding to the F25SrF$_2$ image displays 3 recurrent peaks of auroral signal beyond 3$\sigma$ until 14 minutes after the start of the exposure. These peaks are statistically significant, as a random gaussian distribution of the same number of points shall result in 0.23 and 0.27 data points respectively with an amplitude in excess of 3$\sigma$ above the mean level. Taken altogether, these results give evidence that the auroral region was active during at least 36~min, which increases our above first, rough, 13~min estimate. A closer inspection of the right-handed histograms, which were built from the brightest Uranus auroral emission ever seen with HST (see table \ref{tab}), provides further information on the auroral short-term dynamics. The 10~s integrated histogram shows 3 auroral bursts above 3$\sigma$ and 3 more reaching 2$\sigma$, which repeat along the interval, spaced by several minutes. These bursts are brief and made of individual pulses lasting for less than 1-2~s. The 1~s integrated histogram for instance displays 15 pulses at or in excess of the 3$\sigma$ level (while a gaussian distribution predicts that only 2.7, hence 3 data points shall randomly reach this level) and many more at the 2$\sigma$ level. The Fourier transform of the 1~s integrated histogram (not shown) displays several peaks of moderate amplitude, the most intense one being at 2.5~min (secondary peaks are visible at 0.1, 0.45 and 1.3~min). This 2.5~min recurrence is tentatively indicated with double arrows on the 10~s integrated histogram. While the reliability of this quasi-period deserves to be confirmed over a more statistical dataset, it is interesting to note that similar quasi-periodic polar auroral flares with timescales of several minutes, attributed to dayside pulsed reconnection, have similarly been observed at Earth and Jupiter \citep[and refs therein]{Bonfond_GRL_11}. 

\subsection{Localization of magnetic poles}

In Figure \ref{fig2}, model southern auroral ovals are displayed in red (the associated blue northern ovals are not visible as they are located on the nightside). They were derived from the most up-to-date AH5 magnetic field model of Uranus \citep{Herbert_JGR_09} and delimited by a pair of solid lines which map the footprints of magnetic field lines whose apex reaches 5 (outer line) and 20 (inner line) Uranian radii respectively (1 R$_U$ = 25559~km) at the 1100~km altitude. This wide interval provides a fair guide to investigate any auroral field lines, as it encompasses most of the inner magnetosphere (the 1986 aurorae lay at the footprint of AH5 field lines of apex just outside 5~R$_U$) and the outer magnetosphere (the sub-solar standoff distance of the magnetopause lay at 18~R$_U$ during the Voyager 2 flyby, and is likely to be less during magnetospheric compressions).


In order to quantitatively constrain the longitude of the magnetic poles, we have built a composite cylindrical brightness map from all the 2014 images, including those which did not exhibit any significant auroral signal to take into account any possible weak or diffuse additional aurorae not investigated above. The result is displayed on figure \ref{fig3}. As a result of the planetary inclination, the projection maps all longitudes, and latitudes $\le50^\circ$. We then built a mask from model auroral ovals defined above, and performed a 2D cross-correlation between the two projections by shifting the mask in longitude. This assumes that the latitude of magnetic poles had not varied since 1986. The correlation coefficient clearly peaks twice at 0.15 and 0.13, above an average level of 0.05, for longitudes of the southern magnetic pole of 104$^\circ$ and 118$^\circ$, respectively. We chose the first peak as best fit, and used it to fix the longitude of both magnetic poles. The corresponding model ovals are overplotted on the data in figure \ref{fig3}. The existence of a second peak of comparable (although lower) amplitude simply illustrates that the aurorae, mainly clustered around one localized active region, cannot be uniquely fitted : the oval corresponding to the second fit is located to the right on figure \ref{fig3}. The half maximum of the highest correlation coefficient yields a conservatively acceptable range of 78-130$^\circ$ longitude. Therefore, we identify the southern (northern) magnetic pole at $104\pm26^\circ$ (284$\pm26^\circ$) longitude over the month of Nov. 2014. The subsequent update of the rotation period and ULS system using the full set of HST auroral detections is beyond the scope of this paper.

A similar approach could not be applied to the 2012 observations, because of less frequent and weaker auroral emissions. The model ovals displayed in figures \ref{fig2}a$_2$-b$_2$ thus simply indicate a visual best fit.

\section{Discussion}
\label{discus}

The six detections acquired from the 2012 and 2014 HST campaigns now add to the three auroral signatures detected during the 1998 and 2011 HST campaigns. Although the statistics remain limited, this collection nonetheless provides a basis to further investigate possible origins for the observed auroral precipitations. 

The ring-like faint emissions of 1998 were discussed by L12 who proposed that they are powered by some magnetospheric acceleration process, active for an intermediate Solstice-to-Equinox configuration, and able to operate over a wide range of longitudes. This is consistent with the particularly quiet solar wind conditions which prevailed for more than 5 days on both sides of the observations (Figure \ref{fig1}a).

From the persistent localized and dynamic nature of auroral spots observed over the 2011-2014 period on the sunlit hemisphere, post-Equinox Uranus aurorae are a good candidate for cusp emission (as observed at the Earth, Jupiter and Saturn) at or near the boundary between open and closed field lines. The detected aurorae are brief, second-long events, modulated on timescales of minutes and lasting several tens of minutes. L12 already proposed that the 2011 auroral spots could result from impulsive plasma injections through dayside reconnection with the interplanetary magnetic field, expected to be favored once per rotation according to the variable solar wind/magnetosphere geometry.  Interestingly, the 2011 and 2014 auroral features were in each case radiated by a region which, although activated several weeks apart, remained strikingly fixed in latitude and longitude. If we assume that the aurorae are related to dayside reconnection, a fixed emission locus would therefore suggest a stable reconnection site, in contrast with the expectations of \citet{Masters_JGR_14}. We note, however, that such a mapping is generally poorly reliable due to the complex topology of magnetic field lines at the magnetospheric cusps. Furthermore, \citet{Cowley_JGR_13} pointed out that the topology of magnetic field lines wound around the planet by the rotation is likely to be complex and may even prevent dayside reconnection part of the time. Whether injections are triggered by dayside or nightside reconnections cannot be inferred without a better knowledge of the planetary field geometry. 



Further information on any influence of the solar wind is provided by figure \ref{fig1}, which indicates all the HST detections with gray arrows plotted over the interplanetary dynamic pressure, where the size of the arrow is qualitatively proportional to the signal strength. Despite the large $\sim\pm3$~days uncertainty in the arrival time, this global view draws general trends. We first note that the 2014, 2011 (and even 2012) positive detections match episodes of globally enhanced solar wind activity - as consistently predicted by the different MHD models - lasting for several days and made of successive individual pressure fronts. The most intense Uranus aurorae ever observed (24~kR, 8.8~GW) interestingly match a high-pressure episode (P$\ge$0.017nPa for 2 models over 3), the largest ever sampled at Uranus. While the solar wind is known as a driver for part of planetary aurorae in general, it is worth noting that terrestrial cusp aurorae brighten in particular during magnetospheric compressions, their location being controlled by the interplanetary magnetic field orientation \citep{Farrugia_JGR_95}. Possible Uranian cusp aurorae discussed above might thus be similarly triggered by solar wind compressions.Ê

On the other hand, the limited number of positive detections over all the HST observations which sampled long-lasting periods of active solar wind suggests that the Uranus aurorae also likely depend on the planetary field geometry, and therefore on the planetary rotation, as the mean interplanetary magnetic field at 19~AU remains almost entirely azimuthal.



\section{Conclusion}

In this article, we analyzed two HST/STIS imaging campaigns of Uranus acquired in 2012 and 2014 with different temporal coverage under variable solar wind conditions. Their analysis yielded the identification of six additional detections of Uranus' aurorae acquired on 27 Sept. 2012, 1-2 and 24 Nov. 2014. The persistence of auroral signal on consecutive images at the same coordinates provides direct evidence of a rotational motion with the planet. The aurorae were localized from $-50^\circ$ (in 2012) to $-40^\circ$ (in 2014) southern latitudes. The auroral regions of 1-2 and 24 Nov. 2014 were also rotationally phased, which suggests that the same portion of any auroral oval was activated 22 days apart, as in 2011. The detected emissions lasted for tens of minutes. The auroral region of 24 Nov. 2014 was active for at least 36~min and composed of brief pulses of emission, lasting for less than 1-2s and variable on timescales of minutes, with a main recurrence period of $\sim$2.5~min.

The auroral spots radiated 5-24~kR and 1.3-8.8~GW, which are comparable to the intensity of Uranian aurorae observed previously and demonstrate that these can be routinely observed with HST (the four investigated campaigns each included at least one detection). The Nov. 2014 observations were fitted with model auroral ovals which constrained the longitude of the southern (northern resp.) magnetic pole to $104\pm26^\circ$ ($284\pm26^\circ$ resp.) ULS. We suggest that near-equinoctial Uranus aurorae might be pulsed cusp emissions formed by either dayside or nightside reconnection. The time (and possible amplitude) correlation between aurorae and sudden increases of solar wind dynamic pressure may suggest a prominent influence of the solar wind for driving auroral precipitation (to be confirmed), in addition to the planetary field geometry. These results form a basis for further modeling work of magnetic reconnection or full solar wind/magnetosphere interaction using realistic solar wind parameters prevailing during the investigated observations.

The comparative analysis of Uranus' aurorae detected by HST over 16 years shows an overall variation of Uranus auroral properties from a Solstice-to-Equinox situation (1998) to a configuration gradually moving away from Equinox (2011 to 2014). It is essential to pursue observing Uranian aurorae with HST, the most powerful FUV telescope in activity, as the intermediate Equinox-to-Solstice configuration will be reached in 2017. This configuration will provide an opportunity to check the single auroral detection of 1998 under various solar wind conditions and identify the associated magnetospheric dynamics. Neptune, which forms the family of ice giants planets with Uranus, also represents a worthy unexplored target whose aurorae are likely accessible to HST sensitivity. Neptune's magnetosphere is less tilted with denser and longer plasma residence times, and may thus respond to the solar wind in a similar fashion as Uranus does.


\appendix
\section{Solar wind propagation models}
\label{app_sw}

\subsection{mSWiM} 

The mSWiM 1D model considers the solar wind as an ideal MHD fluid propagated from spacecraft in situ measurements at 1~AU outward in the solar system in a spherically symmetric configuration. The model was originally developed and extensively validated for propagation to between 1 and 10~AU \citep{Zieger_JGR_08} (1 AU = 1 astronomical unit). The input boundary conditions at 1~AU are rotated to an inertial longitude. Propagation occurs at the inertial longitude and then results are rotated to the target body.  Motion of the both the spacecraft providing the boundary conditions and the target body are taken into account. As expected, the model provides the most accurate results when the sun-spacecraft-target are aligned in heliographic longitude. Both the L12 study and the present one use a modified version of this code where the mass loading due to interstellar neutrals in the outer heliosphere (10-20~AU) is taken into account.


\begin{figure*}
\centering
\noindent\includegraphics[width=28pc]{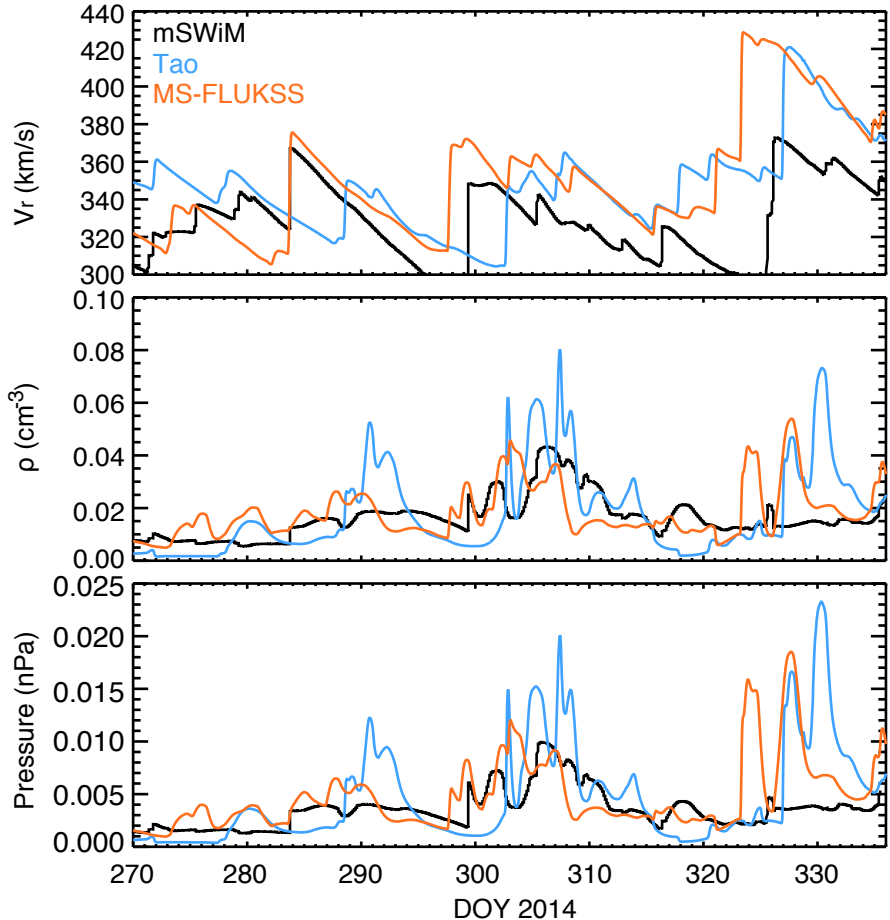}
\caption{Solar wind velocity, density and dynamic pressure predicted at Uranus by the mSWiM (black), Tao (blue) and MS-FLUKSS (orange) models for late 2014.}
\label{figS1}
\end{figure*}

\subsection{Tao model} 

The Tao 1D model considers the solar wind as an ideal MHD fluid in a one-dimensional spherical symmetric coordinate system. The equation set, numerical scheme, model setting, and inputs are detailed in \citep{Tao_JGR_05}. The modifications brought to the code to propagate solar wind up to the Uranus orbit are described below.

To account for the effect of the solar rotation, the solar wind arrival time is delayed by $\Delta t =\Delta\Phi/\Omega$, where $\Delta\Phi$ is the Earth-Sun-Uranus angle and $\Omega$ is the solar angular velocity (using a 26 days rotation period). 

In the outer heliosphere (beyond 10~AU), the interaction between the solar wind and the neutral hydrogen of the local interstellar medium becomes non-negligible. It is taken into account by assuming that the neutral hydrogen distribution and the temperature vary as a function of the heliospheric distance $r$ as follows. 

The hydrogen density $n_H(r)$ and velocity $u_H(r)$ are defined as in \citep{Wang_JGR_01} (equation 7) : $n_H (r)=n_H^{\infty}\exp^{-\lambda/r}$ and $u_H (r)=u_H^{\infty}$ with $\lambda= 7.5$~AU, $n_H^{\infty}=0.09$~cm$^{-3}$ \citep{Wang_JGR_03} and $u_H^{\infty}=20$~km/s. The direction of the interstellar wind is used to derive the radial and azimuthal components of the velocity along the the Sun-Uranus reference line \citep{Lallement_AIP_10}. 

The temperature profile is defined as in \citep{Wang_JGR_03} : $T_H(r)=1000+T_H^{\infty}\exp^{-\lambda/r}$, where $T_H^{\infty}= 1.09\times10^4$~K.

The interaction of the solar wind with the neutral hydrogen is introduced through the momentum and energy equations following the description of \citet{Mcnutt_JGR_98} (see equations 29, 70 and 71). The energy source term is multiplied by 1.8 in order to obtain a steady state proton temperature profile consistent with Voyager 2 observations (e.g. Figure 1 of \citep{Wang_JGR_03}).

\subsection{MS-FLUKSS} 

\citet{Kim_ApJ_16} recently developed a 3D model which predicts solar wind conditions between 1 and 80~AU from time-dependent boundary conditions implemented in the adaptive mesh refinement framework of Multi-scale Fluid-kinetic Simulation Suite (MS-FLUKSS), which is a numerical toolkit designed primarily for modeling flows of partially ionized plasma (see \citep[and refs therein]{Pogorelov_14}). MS-FLUKSS solves MHD equations for plasma coupled either with the kinetic Boltzmann or multiple gas dynamics Euler equations describing the flow of different populations of neutral atoms. Several different turbulence models are implemented in MS-FLUKSS together with different approaches to treat non-thermal (pickup) ions as a separate plasma components. In this particular simulation, the model takes into account the effects of pickup ions that are created in the charge-exchange process between the solar wind and interstellar neutral atoms. While the flows of plasma and neutral atoms are described separately by solving the MHD and Euler equations, respectively, the thermal (solar wind) and non-thermal (pickup ions) plasma are treated as a single, isotropic fluid. Thus, the model plasma temperatures are generally greater than those expected for the solar wind at distances greater than $\sim$10~ AU such as at Uranus, due to the contribution from the much hotter pickup ions that become increasingly dominant at larger distances. 

\subsection{Comparison of the model predictions at Uranus}

As only the MS-FLUKKS results have been validated yet in the outer heliosphere, these results are hereafter taken as a reference to which the mSWiM and Tao results are compared to assess typical uncertainties.

The solar wind parameters at Uranus predicted by these three models are compared on Figure \ref{figS1} throughout a representative time interval of 66~days , which encompasses the Nov. 2014 HST observations. The most accurately propagated parameters are the radial velocity (top panel) and the density (middle panel), or their combination within the dynamic pressure (bottom panel), whose sudden increases indicate interplanetary shocks. Results from MS-FLUKKS, mSWiM and the Tao model and MS-FLUKKS are displayed in orange, black and blue, respectively. 

Figure \ref{figS1} illustrates a general agreement between the results of the three models which all predict three different disturbed solar wind episodes separated by three quiet conditions episodes. We note that mSWiM's densities are generally lower than those of MS-FLUKKS and Tao. In addition, these densities remain strikingly low and constant after DOY 320, while the mSWiM's densities calculated without considering interstellar neutrals (not shown) are more consistent with MS-FLUKKS's and Tao's ones during this period. The mSWiM's predictions are thus considered as insufficiently reliable after day 320 of year 2014. 

The delay between the arrival of velocity, density or pressure fronts predicted by the three models varies from 1 to 5 days, from 2 to 5 days and from 2 to 4 days during the three active solar wind periods (DOY 284-292, 298-309 and 323-331, respectively). Consequently, we have set an estimate of $\pm3$~days uncertainty, as indicated in the main text. However, many individual fronts apparent in Figure S1 (late 2014) and most of the fronts displayed in Figure 1 (mid 1998, late 2011, late 2012) display a much better coincidence.

\begin{acknowledgments}
This work is based on observations of the NASA/ESA Hubble Space Telescope (Programs GO \#13012 and GO/DD \#14036). We thank the Space Telescope Science Institute staff for their support in scheduling the observations. The French co-authors acknowledge support from CNES and thank L. Gosset who worked on a preliminary analysis of the 2014 data during a master internship in 2015. We acknowledge the APIS service http://apis.obspm.fr, hosted by the Paris Astronomical Data Centre, and regularly used to browse and investigate the data, and thank in particular F. Henry and P. Le Sidaner who maintain this service. We acknowledge use of NASA/GSFC's Space Physics Data Facility's OMNIWeb service http://omniweb.gsfc.nasa.gov and OMNI data. We thank the CCMC and L. Mays to have ran the ENLIL model specifically at Uranus for comparison purposes. Work at Leicester was supported by STFC grant ST/N000749/1. SVB was supported by STFC Fellowship ST/M005534/1. TK and NP were supported by NASA grants NNX14AF41G and NNX14AJ53G, and NSF SHINE grant AGS-1358386.
\end{acknowledgments}

\end{article}

\end{document}